\documentclass[12pt]{iopart}
\usepackage{graphicx}
\usepackage{epsfig}
\usepackage{epstopdf}
\usepackage{color}
\usepackage[normalem]{ulem}
\usepackage{bm}

\begin{document}

\title{Calculation of Berry curvature using nonorthogonal atomic orbitals}

\author{Jin Gan, Daye Zheng, Lixin He*}
\address{Key Laboratory of Quantum Information, University of Science and
  Technology of China, Hefei, Anhui, 230026, People's Republic of China}
\address{Synergetic Innovation Center of Quantum Information and Quantum
  Physics, University of Science and Technology of China, Hefei, 230026, China}
\ead{helx@ustc.edu.cn}

\begin{abstract}

We present a derivation of the full formula to calculate the Berry curvature on non-orthogonal numerical
atomic orbital (NAO) bases.
Because usually, the number of NAOs is larger than that of the Wannier bases, we use a orbital contraction
method to reduce the basis sizes, which can greatly improve the calculation efficiency
without significantly reducing the calculation accuracy.
We benchmark the formula by calculating the Berry curvature of ferroelectric BaTiO$_3$ and bcc Fe,
as well as the anomalous Hall conductivity (AHC) for Fe. The results are in excellent agreement with the finite difference and previous results in the literature. We find that there are corrections terms to the Kubo formula of the Berry curvature.
For the full NAO base, the differences between the two methods are negligibly small, but for the reduced bases sets, the correction terms become larger, which may not be neglected in some cases.
The formula developed in this work can readily be applied to the non-orthogonal generalized Wannier functions.
\end{abstract}
\maketitle

\section{Introduction}

Berry curvature is of fundamental importance for understanding some basic properties of
solid materials and is essential for the description of the dynamics of
Bloch electrons~\cite{Di2010,vanderbilt_2018}.
It acts as a magnetic field in momentum space, which leads to some anomalous transport effects,
including the first order\cite{Jungwirth2002,Onoda2002,Fang2003,Yao2004,Nagaosa2010,Di2010,Weng2015},
and second order\cite{Sodemann2015,Zhang20181,Zhang20187,You2018,Ma2019} anomalous Hall effects (AHE).
It may introduce shift current in polar materials \cite{Sipe2000,Young2012,Tan2016,Wang2019}.
Berry curvature also plays a crucial role in the classification of topological materials
\cite{Hasan2010,Qi2011,Kane2005}.

The first-principles calculations of the Berry curvature
and the anomalous Hall conductivity (AHC) have been reviewed in Ref.~\cite{Gradhand2012}.
Especially, the Berry curvature and AHC have been calculated via the Kubo formula \cite{Fang2003,Yao2004}.
However, this method requires to calculate band structures at very dense $k$ points, and sum over a large number of unoccupied states to converge the results. Therefore the computational cost is very high, even for simple materials.
Vanderbilt and co-works developed a very efficient interpolation scheme \cite{Wang2006} to calculate Berry curvature and AHC \cite{Sundaram1999,Adams1959} based on maximally
localized Wannier functions (MLWFs)\cite{Marzari1997,Souza2001}, which has been demonstrated to calculate the AHC to very high accuracy with only a tiny fraction of time of the original Kubo formula.
However, it is not always easy to construct high-quality MLWFs, for complex systems. More seriously, in some cases, the MLWFs may not respect the point group symmetries of the crystal. 
Breaking of the symmetry may lead to qualitatively incorrect results. Special attention has to be paid to construct the symmetry adapted MLWFs \cite{Souza2001,Sakuma2013,Thygesen2005}.

Recently, Lee at. el. derived the Kubo formula of optical matrices as well as Berry curvature
using nonorthogonal numerical atomic orbitals (NAOs)~\cite{Lee2018}. Wang et. al. also derived the formula of Berry connection and its higher order derivatives based on NAOs~\cite{Wang2019}. The NAOs are strictly localized and more importantly, have spherical symmetries, and
therefore, no extra effort is needed to construct the symmetry-adapted MLWFs. This is of great advantage for the applications in complex systems.
In this work, we derive the full formula to calculate the Berry curvature bases on NAOs~\cite{Chen2010}. We show that in the full derivation of Berry curvature, there are correction terms to the Kubo formula\cite{Lee2018}
on the NAO bases.
Since the number of NAOs is usually larger than that of the MLWFs,
we use a orbital contraction technique to reduce the number of NAOs,
which can greatly improve the calculation efficiency without significantly  reducing the calculation accuracy.
These correction terms are small if a large NAO base is used, however, for the reduced NAO bases, the corrections are not negligible in some cases.
The formula developed in this work can be directly applied to the non-orthogonal generalized Wannier functions,
which, if constructed properly, are typically more localized than the orthogonal Wannier functions \cite{he2001}.

The rest of the paper is organized as follows. In Sec.~\ref{sec:formula}, we present the derivation of Berry curvature for non-orthogonal NAOs, followed by detailed benchmark calculations of Berry curvature for BaTiO$_3$ and AHC for Fe in
Sec.~\ref{sec:results}.
In Sec.~\ref{sec:reduce_NAO}, we introduce a technique to reduce the number of NAOs
to accelerate Berry curvature calculations. We summarize in Sec.~\ref{sec:summary}.

\section{Berry curvature in non-orthogonal atomic bases}
\label{sec:formula}

\subsection{Derivation of Berry curvature formula}

The derivation of Berry curvature formula expressed in a non-orthogonal NAO basis
is similar to that of the orthogonal Wannier bases, but there are also some considerable differences.
The derivation is transparent for people who are not familiar with the Wannier functions.

We start from the definition of the Berry curvature \cite{vanderbilt_2018},
\begin{equation}
\mathbf{\Omega}_n(\mathbf{k}) = \nabla \times \mathbf{A}_n(\mathbf{k})\, ,
\end{equation}
where,
\begin{equation}
\mathbf{A}_n = i\langle u_{n\mathbf{k}}|\nabla_{\mathbf{k}}|u_{n\mathbf{k}}\rangle
\end{equation}
is the Berry connection, and $u_{n\mathbf{k}}$ are the cell-periodic Bloch functions,
whose expression on a NAO basis is given by Eq.~(\ref{eq:u_eq}) in the Appendix.

More generally, 
one may generalize the above
definition of Berry connection and Berry curvature to a multi-bands case,
\begin{equation}
A_{n m, \alpha}(\mathbf{k})=i\left\langle u_{n\mathbf{k}} \mid \partial_{\alpha} u_{m\mathbf{k}}\right\rangle\, ,
\label{eq:berryconnection}
\end{equation}
and
\begin{eqnarray}
{\Omega}_{nm,\alpha\beta}(\mathbf{k})
&=& \partial_\alpha\mathbf{A}_{nm,\beta}(\mathbf{k}) - \partial_\beta\mathbf{A}_{nm,\alpha}(\mathbf{k}) \nonumber \\
&=& i\langle\partial_\alpha u_{n\mathbf{k}}|\partial_\beta u_{m\mathbf{k}}\rangle - i\langle\partial_\beta u_{n\mathbf{k}}|\partial_\alpha u_{m\mathbf{k}}\rangle\, ,
\label{eq:omega_m}
\end{eqnarray}
where $\partial_{\alpha}=\partial / \partial k_{\alpha}$.

Substituting the cell-periodic wave functions Eq.~(\ref{eq:u_eq}) into Eq.~(\ref{eq:omega_m}), we have,
\begin{eqnarray}
& &i\langle\partial_\alpha u_{n\mathbf{k}}|\partial_\beta u_{m\mathbf{k}}\rangle\nonumber \\
&=&i\sum_{\nu,\mu}C_{n\nu}^{*}C_{m\mu}\sum_{\vec{R}}\mathrm{e}^{i\mathbf{k}\cdot\mathbf{R}}\langle\mathbf{0}\nu|-r_\alpha(R_\beta-r_\beta)|\mathbf{R}\mu\rangle \nonumber \\
&+& i\sum_{\nu,\mu}\left(\partial_\alpha C_{n\nu}^{*}\right)S_{\nu\mu}\left(\partial_\beta C_{m\mu}\right) \nonumber \\
&+& \sum_{\nu,\mu}\left(\partial_\alpha C_{n\nu}^{*}\right)C_{m\mu}\sum_{\mathbf{R}}\mathrm{e}^{i\mathbf{k}\cdot\mathbf{R}}\langle0\nu|r_\beta-R_\beta|\mathbf{R}\mu\rangle \nonumber \\
&-& \sum_{\nu,\mu}C_{n\nu}^{*}\left(\partial_\beta C_{m\mu}\right)\sum_{\mathbf{R}}\mathrm{e}^{i\mathbf{k}\cdot\mathbf{R}}\langle0\nu|r_\alpha|\mathbf{R}\mu\rangle \label{uu_eq}
\end{eqnarray}

To simplify the above equation, we introduce a dipole matrix $A^R_\alpha$, as follows,
\begin{equation}\label{eq:dipole}
A^R_{\nu\mu,\alpha}(\mathbf{k}) = \sum_{\mathbf{R}}\mathrm{e}^{i\mathbf{k}\cdot\mathbf{R}}\langle\mathbf{0}\nu|r_\alpha|\mathbf{R}\mu\rangle\, ,
\end{equation}
where the superscript ``$R$'' in $A^R$ refers to that the dipole matrix is summed over the lattice ${\bf R}$.
This quantity is similar to the $A^{W}_{\nu\mu,\alpha}$ in Ref.~\cite{Wang2006}.
While it is somehow cumbersome to calculate the dipole matrix in the Wannier bases,
$A^R_{\nu\mu,\alpha}$ can be easily calculated by two-center integrals on the NAO bases
by taking the advantages of the spherical symmetry of the NAOs\cite{LI2016503,siesta-ref}.

One part of the contribution to the Berry curvature is from the dipole matrix,
\begin{equation}
\bar{A}_{nm,\alpha} = C_n^\dagger A^R_\alpha C_m \, ,
\end{equation}
which is due to the lack of inversion symmetry of the crystal.

The other contribution to the Berry curvature comes from the change of the Bloch wave function
coefficient $C_{n}({\bf k})$ with ${\bf k}$.
Following Ref.~\cite{Wang2006}, we may also introduce a $\mathbf{D}$ matrix in the non-orthogonal NAO bases, as follows,
\begin{equation}
D_{nm,\alpha} = C_n^\dagger S\left(\partial_\alpha C_m\right) \, .
\end{equation}
There are useful relations between $\mathbf{\bar{A}}$ and $\mathbf{\bar{A}^\dagger}$, $\mathbf{D}$ and $\mathbf{D}^\dagger$, and the proof is given in the Appendix.
\begin{eqnarray}
\bar{A}_{nm,\alpha} - (\bar{A}^\dagger)_{nm,\alpha} &=& -i\bar{S}_{nm,\alpha} \label{eq:relation1}\\
D_{nm,\alpha} + (D^\dagger)_{nm,\alpha} &=& -\bar{S}_{nm,\alpha} \label{eq:relation2}
\end{eqnarray}
where
\begin{equation}
\bar{S}_{nm,\alpha} = C_n^\dagger\left(\partial_\alpha S\right)C_m \, .
\end{equation}
For the orthogonal Wannier bases, where $\bar{S}_{nm,\alpha}$=0, we have $\bar{A}_{nm,\alpha} = (\bar{A}^\dagger)_{nm,\alpha}$,
and $D_{nm,\alpha} = -(D^\dagger)_{nm,\alpha}$, i.e,  $\bar{\bf A}$ is Hermitian, and ${\bf D}$ is anti-Hermitian
in the orthogonal Wannier bases.
It is easy to show that Berry connection $\mathbf{A}_{mn}=i{\bf D}_{mn}+{\bf \bar{A}}^\dagger_{mn}$.

We can simplify Eq.~(\ref{uu_eq}) by inserting the identity matrix,
\begin{equation}
I = \sum_{n}C_{n}C_{n}^\dagger S = \sum_{n}SC_nC_n^\dagger \, .
\end{equation}
For example, for the second term on the right side of Eq.~(\ref{uu_eq}), we have,
\begin{eqnarray}
& i\sum_{\nu,\mu}\left(\partial_\alpha C_{n\nu}^{*}\right)S_{\nu\mu}\left(\partial_\beta C_{m\mu}\right) \nonumber\\
&= i\left(\partial_\alpha C_n^\dagger\right)S\left(\partial_\beta C_m\right) \nonumber \\
&= i\sum_{l}\left(\partial_\alpha C_n^\dagger\right)SC_lC_l^\dagger S\left(\partial_\beta C_m\right) \nonumber \\
&= i\sum_{l}(D^{\dagger})_{nl,\alpha} D_{lm,\beta}\, .
\end{eqnarray}

After some derivation, we obtain the formula of the Berry curvature in the non-orthogonal NAO bases,
\begin{eqnarray}
\Omega_{nm,\alpha\beta} &= \bar{\Omega}_{nm,\alpha\beta} + i\left(D^\dagger_\alpha D_\beta-D^\dagger_\beta D_\alpha\right)_{nm} \nonumber \\
& +\left(D^\dagger_\alpha\bar{A}^\dagger_\beta+\bar{A}_\beta  D_\alpha\right)_{nm} -\left(D^\dagger_\beta\bar{A}^\dagger_\alpha+\bar{A}_\alpha D_\beta\right)_{nm} \, ,
\label{eq:origin_curv}
\end{eqnarray}
where
\begin{equation}
\label{eq:omega_bar}
\bar{\Omega}_{nm,\alpha\beta} = i\sum_{\nu,\mu}C_{n\nu}^{*}C_{m\mu}\sum_{\mathbf{R}}\mathrm{e}^{i\mathbf{k}\cdot\mathbf{R}}\langle0\nu|r_\beta R_\alpha-r_\alpha R_\beta|\mathbf{R}\mu\rangle \, .
\end{equation}
We note that Eq.~(\ref{eq:origin_curv}) is very similar to Eq.~(27) in Ref.~\cite{Wang2006}, but also with notable differences. Because $\bar{\bf A}$ is not Hermitian, and ${\bf D}$ is not anti-Hermitian for the non-orthogonal NAO bases, we cannot write the equation in the form of commutators as Eq.~(27) in Ref.~\cite{Wang2006}.

\subsection{Calculation of $\mathbf{D}$ matrix}
\label{sec:IIB}

From the linear response theory given in Appendix A, one obtains\cite{Wang2019} (for $m \neq n$),
\begin{equation}
D_{nm,\alpha} =
\frac{\bar{H}_{nm,\alpha}-E_{m\mathbf{k}} \bar{S}_{nm,\alpha}}{E_{m\mathbf{k}}-E_{n\mathbf{k}}} \, ,
\label{eq:d-matrix}
\end{equation}
where
\begin{equation}
\bar{H}_{nm,\alpha} = C_n^\dagger\left(\partial_\alpha H\right)C_m \, .
\end{equation}
There is some freedom to choose ${\bf D}_{nn}$.
However, since ${\bf D}_{nn}=  C_n^\dagger ({\bf k}) S({\bf k})\partial_{\bf k}  C_n({\bf k})$,
it must satisfy the following constrain,
\begin{equation}
{\bf D}_{nn}^\dagger + {\bf D}_{nn}= -C_n^\dagger ({\bf k})\partial_{\bf k} S({\bf k})\, C_n({\bf k}) \, .
\label{eq:Dnn}
\end{equation}
For the orthogonal bases, one can just take ${\bf{D}}_{nn}=0$ \cite{Wang2006}.
However, this choice is generally not feasible for the nonorthogonal base.
Instead, we can use the parallel transport gauge,
$\mathbf{A}_{nn}=i\left\langle u_{n}| \partial_{\bf k} u_{n}\right\rangle=0$, i.e.,
${\bf D}_{nn}=i{\bf \bar{A}}^\dagger_{nn}$.
Using the relation, $ \mathbf{A}_{mn}=i{\bf D}_{mn}+{\bf \bar{A}}^\dagger_{mn}$,
and the relations in Eq.~(\ref{eq:relation1}) and Eq.~(\ref{eq:relation2}) for $m$=$n$,
we can easily prove that the parallel transport gauge, automatically satisfies
the constrain of Eq.~(\ref{eq:Dnn}). Note that the parallel transport gauge
defined here for the cell-periodic functions
is different from the ``parallel transport gauge''
in Ref.\cite{Wang2006} for the state vectors $||\phi_n \rangle\rangle $
in the ``tight-binding space'' of the MLWFs.
As shown in the next section, ${\bf D}_{nn}$ would not appear in the Berry curvature calculations, and therefore the
choice of particular gauge would not change the Berry curvature, but
it may affect the Berry connection.

\subsection{Total Berry curvature}

The total Berry curvature is calculated as follows,
\begin{equation}
\Omega_{\alpha\beta}(\mathbf{k})= \sum_{n}f_n(\mathbf{k})\Omega_{nn,\alpha\beta}(\mathbf{k})\, ,
\end{equation}
where, $f_n$ is the Fermi occupation function.
To avoid the numerical instability caused
by the canceling contributions of large values
of $D_{nm}$ [see Eq.~(\ref{eq:d-matrix})],
originated from the small energy splitting between
a pair of occupied bands $n$ and $m$,
we would like to
sperate the summation between the occupied and unoccupied states,
similar to the MLWF interpolation method\cite{Wang2006}.
However, this is a little bit more tricky for the non-orthogonal bases.

We first rewrite Eq. (\ref{eq:origin_curv}) by replacing $D^\dagger_\alpha$ and $D^\dagger_\beta$
with $D_\alpha$ and $D_\beta$  using Eq.~(\ref{eq:relation2}),
\begin{eqnarray}\label{eq:modify_omega}
\Omega_{\alpha\beta}&=& \bar{\Omega}_{\alpha\beta} - \left[D_\alpha, \bar{A}^\dagger_\beta\right]
+\left[D_\beta,\bar{A}^\dagger_\alpha\right]\nonumber \\
&- &i\left[D_\alpha, D_\beta \right]
- \left(\bar{S}_\alpha\bar{A}^\dagger_\beta - \bar{S}_\beta\bar{A}^\dagger_\alpha\right)\, ,
\end{eqnarray}
where the first four terms closely resemble those of Eq.~(27) in Ref.\cite{Wang2006}.
One can proof the Berry curvature defined above is gauge invariant (see Appendix A3).
In this form, the contribution from $D$ matrices are exactly canceled for a pair of occupied
states $n$, $m$. The last term is due to the non-orthogonality of the NAO bases.
We can therefore calculate the total Berry curvature $\Omega_{\alpha\beta}$ as~\cite{Wang2006},
\begin{eqnarray}\label{eq:total_berrycurvature}
\Omega_{\alpha\beta}(\mathbf{k})
&=& \sum_{n}f_n\bar{\Omega}_{nn,\alpha\beta} \nonumber
+ \sum_{n,m}(f_m-f_n) [ i D_{nm,\alpha}D_{mn,\beta}  \nonumber \\
&+& D_{nm,\alpha}(\bar{A}^\dagger)_{mn,\beta} - D_{nm,\beta}(\bar{A}^\dagger)_{mn,\alpha} ] \nonumber \\
&-& \sum_{n,m}f_n\left[  \bar{S}_{nm,\alpha}(\bar{A}^\dagger)_{mn,\beta}
- \bar{S}_{nm,\beta}(\bar{A}^\dagger)_{mn,\alpha} \right]  \label{Omega_eq}
\end{eqnarray}
We immediately see that for the orthogonal bases, the last line in the above equation vanishes,
and the total Berry curvature reduces to Eq.~(32) in Ref.\cite{Wang2006}.

\subsection{Comparison with the naive Kubo formula}

The Berry curvatures are often calculated via the naive Kubo formula~\cite{Gradhand2012,Lee2018},
\begin{equation}
\label{kuob_eq}
\Omega_{\alpha\beta}^{{\rm kubo}}(\mathbf{k}) = -2\,{\rm Im} \sum_{n}^{occ}\sum_{m}^{uocc}\frac{\upsilon_{nm,\alpha}\upsilon_{mn,\beta}}{(E_{m\mathbf{k}}-E_{n\mathbf{k}})^2}\,,
\end{equation}
where $\upsilon_{nm,\alpha}$ is the velocity matrix.
According to Ref.~\cite{Yates2007,Blount1962305}, we have
\begin{equation}
\upsilon_{nm,\alpha}(\mathbf{k}) 
= \left(\partial_\alpha E_{n\mathbf{k}}\right) \delta_{nm}
- i\left(E_{m\mathbf{k}}-E_{n\mathbf{k}}\right)A_{nm,\alpha}\, ,
\end{equation}
where the Berry connection $A_{nm,\alpha}=iD_{nm,\alpha}+ \bar{A}^\dagger_{nm}$. One easily obtain
the velocity operator matrix for $m$$\neq$$n$,
\begin{equation}
\upsilon_{nm,\alpha}(\mathbf{k})= \bar{H}_{nm,\alpha} - E_{n\mathbf{k}}\bar{S}_{nm,\alpha} + i(E_{n\mathbf{k}}-E_{m\mathbf{k}})\bar{A}_{nm,\alpha} \,,
\end{equation}
which is identical to that derived in Ref.~\cite{Lee2018}
via a different approach.

By comparing Eq.~(\ref{Omega_eq}) and Eq.~(\ref{kuob_eq}), one finds that
the full Berry curvature actually includes some correction terms to the naive Kubo formula, i.e.,
$\Omega_{\alpha\beta}=\Omega_{\alpha\beta}^{{\rm kubo}}+\Delta\Omega_{\alpha\beta}$,
and,
\begin{eqnarray}
 \Delta\Omega_{\alpha\beta}
& = &\sum_{n}^{occ}\bar{\Omega}_{nn,\alpha\beta} - i\sum_{n}^{occ}\sum_{m}^{all} [\bar{A}_{nm,\alpha} (\bar{A}^{\dagger})_{mn,\beta} \nonumber \\
&-& \bar{A}_{nm,\beta} (\bar{A}^\dagger)_{mn,\alpha} ]\, ,
\end{eqnarray}
where $\bar{\Omega}$  is defined in Eq.~(\ref{eq:omega_bar}).
These additional terms are also presented in the MLWF bases~\cite{Wang2006}.
These correction terms come from the incompleteness of
the tight-binding bases to the original Hilbert space~\cite{Graf1995,Boykin1995}.
Even though the contributions of $\Delta\Omega_{\alpha\beta}$ are often very small, and sometimes negligible,
it is still useful to have a strict formula to compare with.
This will be further addressed in the following sections.

\section{Results and discussion}
\label{sec:results}

We take BaTiO$_3$ and ferromagnetic Fe as examples to
benchmark the calculation of the Berry curvature using NAOs.
These choices represent two typical systems where Berry curvature is nonzero that either
break the inversion symmetry (BaTiO$_3$) or time-reversal symmetry (Fe). While the SOC is not needed for BaTiO$_3$
to get nonzero Berry curvature, it is essential for Fe to have nonzero Berry curvature, which is well explained
in Ref.\cite{vanderbilt_2018}

\subsection{Computational details}

We first perform self-consistent density functional theory calculation implemented
in the Atomic-orbtial Based Ab-initio Computation at UStc (ABACUS) code \cite{Chen2010,LI2016503}.
The generalized gradient approximation in the Perdew-Burke-Ernzerhof (PBE) \cite{Perdew1996} form is adopted. We adopt optimized norm-conserving Vanderbilt \cite{Hamann2013}  multi-projector, SG15 pseudopotentials \cite{Schlipf2015,Scherpelz2016}.

For the BaTiO$_3$,  the 5s$^2$5p$^6$5d$^1$6s$^1$ electrons for Ba, the 2s$^2$3p$^6$4s$^2$3d$^2$ electrons for Ti,
and the 2s$^2$2p$^4$ electrons for O are treated as valence electrons.
The cutoff energy for the wave function is set to 60 Ry. A $4\times4\times4$ $k$-mesh is used in the self-consistent calculations. The NAO bases for Ba, Ti, O are 4s2p2d, 4s2p2d1f, 2s2p1d, respectively.

For bcc Fe, the 3s$^2$3p$^6$4s$^2$3d$^6$ electrons are included self-consistently. Spin-orbit interactions are tuned on. The cutoff energy for the wave function is set to 120 Ry. A 16$\times$16$\times$16 $k$-mesh is used in the self-consistent calculations. The NAOs of Fe is 4s2p2d1f.

After the self-consistent calculations, the tight-binding Hamiltonian $H(\bf{R})$
and overlap matrices $S({\bf R})$ in the NAO bases, which are generated during the self-consistent calculations,
are readily outputted and are used for the Berry curvature calculations.

\subsection{Berry curvature of BaTiO$_3$}

\begin{figure}[htbp]
	\centering	
	\includegraphics[width=0.45\textwidth]{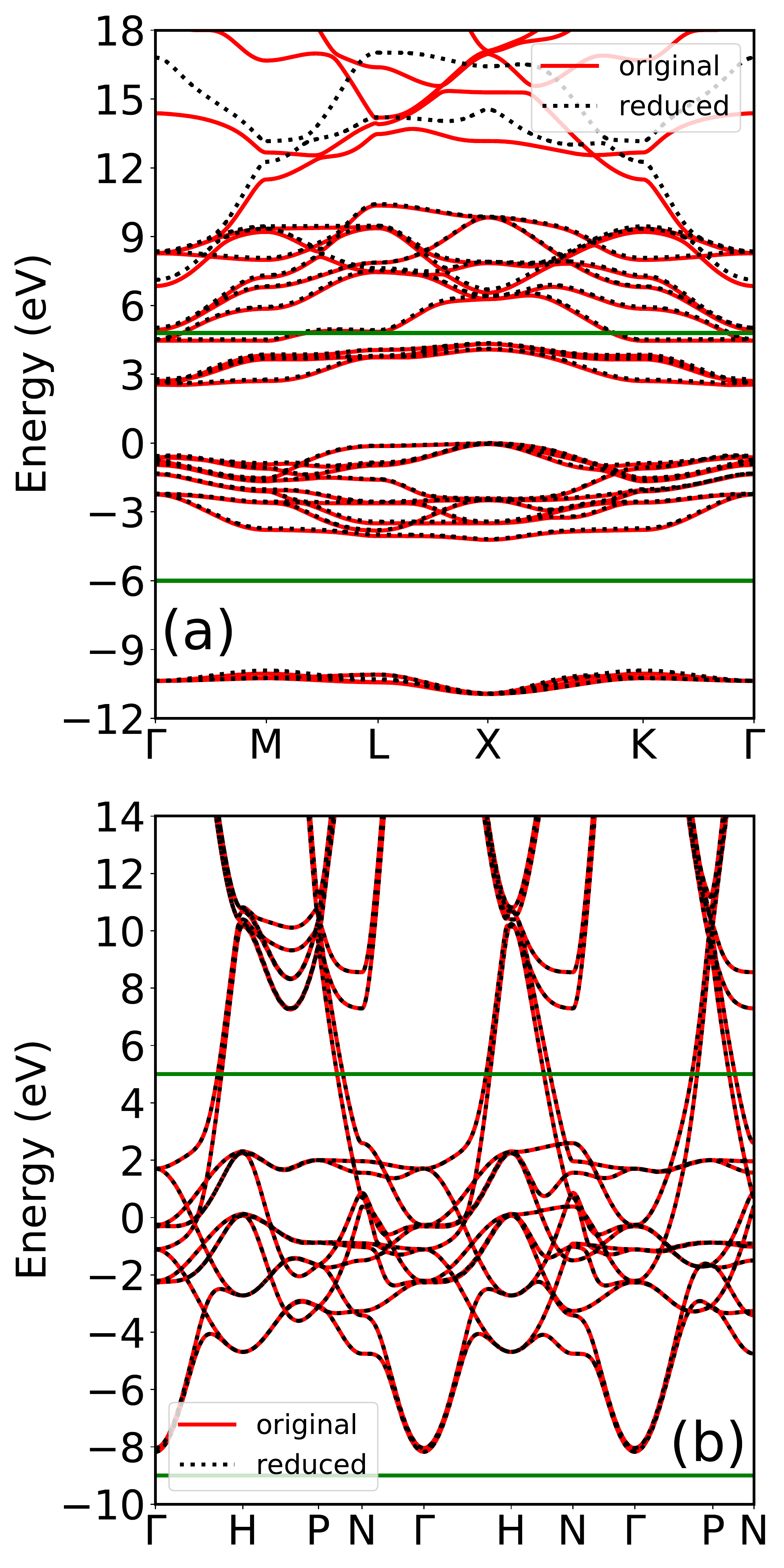}	
	\caption{The band structures of (a) BaTiO$_3$, and (b) bcc Fe. The red solid lines are calculated by the original bases, whereas the black dashed lines are the results of the reduced basis sets. The Fermi levels are set to $E_F$=0, and the green lines indicate the energy windows for the construction of the reduced NAO bases.}
	\label{fig:bands}
\end{figure}

As the first example, we calculate the Berry curvature of rhombohedral BaTiO$_3$ with space group $R3m$ (No.~160, Rhombohedral axes).
The lattice constant $a$ = 4.081 \AA, and $\alpha$ = 89.66$^\circ$ are used. The Wyckoff positions are given in Table \ref{tab:structure_bto}. BTO is a ferroelectric insulator without inversion symmetry.
The band structures of BaTiO$_3$ near the Fermi level $E_F$=0 are shown in red solid lines in Fig.~\ref{fig:bands}(a).

\begin{table}
	\caption{The Wyckoff positions of BaTiO$_3$.}
\center
	\begin{tabular}{c c  c c c}
		\hline
	      & $x$ & $y$ & $z$ & Wyckoff position \\
		\hline
		\hline
		Ba & 0.002 & 0.002 & 0.002 & 1a \\
		Ti & 0.516 & 0.516 & 0.516 & 1a \\
	     O &  0.974 & 0.485 & 0.485 & 3b \\
		\hline
		\hline
	\end{tabular}
	\label{tab:structure_bto}
\end{table}

\begin{figure*}[tbp]
	\centering
\includegraphics[width=0.8\textwidth]{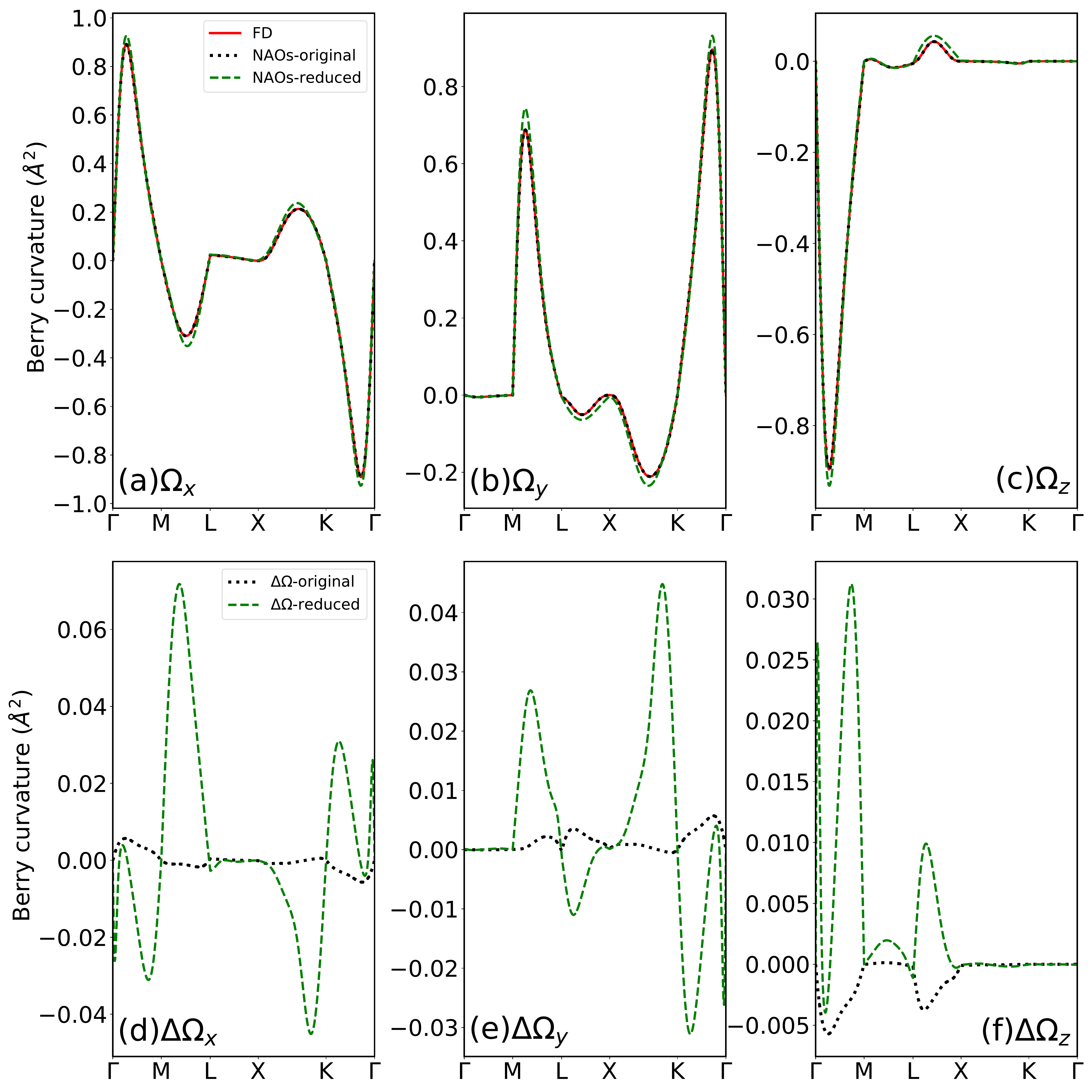}
	\caption{ (Upper panels) The Berry curvatures, (a) $\Omega_x$, (b) $\Omega_y$ and (c) $\Omega_z$ of BaTiO$_3$ along the high symmetry $k$ points.
(Lower panels) The corrections to the Kubo formula, (d) $\Delta \Omega_x$, (b) $\Delta \Omega_y$ and (c) $\Delta \Omega_z$ along the high symmetry $k$ points.The red solid lines are calculated by FD, and the black dotted lines are  results of the original bases, whereas the green dashed lines are the results of the reduced bases.
}
	\label{fig:fig2}
\end{figure*}

Figure \ref{fig:fig2}(a)-(c) depict the Berry curvatures of BaTiO$_3$ along the $a$, $b$ and $c$-axes respectively, where
\begin{equation}
\Omega_{\gamma}(\mathbf{k})=\epsilon_{\alpha \beta \gamma} \Omega_{\alpha \beta}(\mathbf{k}) \, .
\end{equation}
The Berry curvatures calculated by Eq.~(\ref{Omega_eq}) are shown in the black dashed lines. We compared the results to those calculated by the finite-difference (FD) method via Eq.~(\ref{eq:FD}), which are shown in the solid red lines.
As we see, the Berry curvatures calculated by the Eq.~(\ref{Omega_eq}) are in excellent agreement with the FD method.

Figure~\ref{fig:fig2}(d)-(f) show the differences $\Delta\Omega_{\alpha\beta}$ between this work and Kubo formula of BaTiO$_3$
along the $a$, $b$ and $c$-axes respectively, in black solid lines. As seen from the figure, $\Delta\Omega_{\alpha\beta}$ is extremely small, which are less than 1\% of the total Berry curvature. These results suggest that the NAO bases used in the calculations are rather complete for this problem.

\begin{figure}[tbp]
	\centering
	\includegraphics[width=0.45\textwidth]{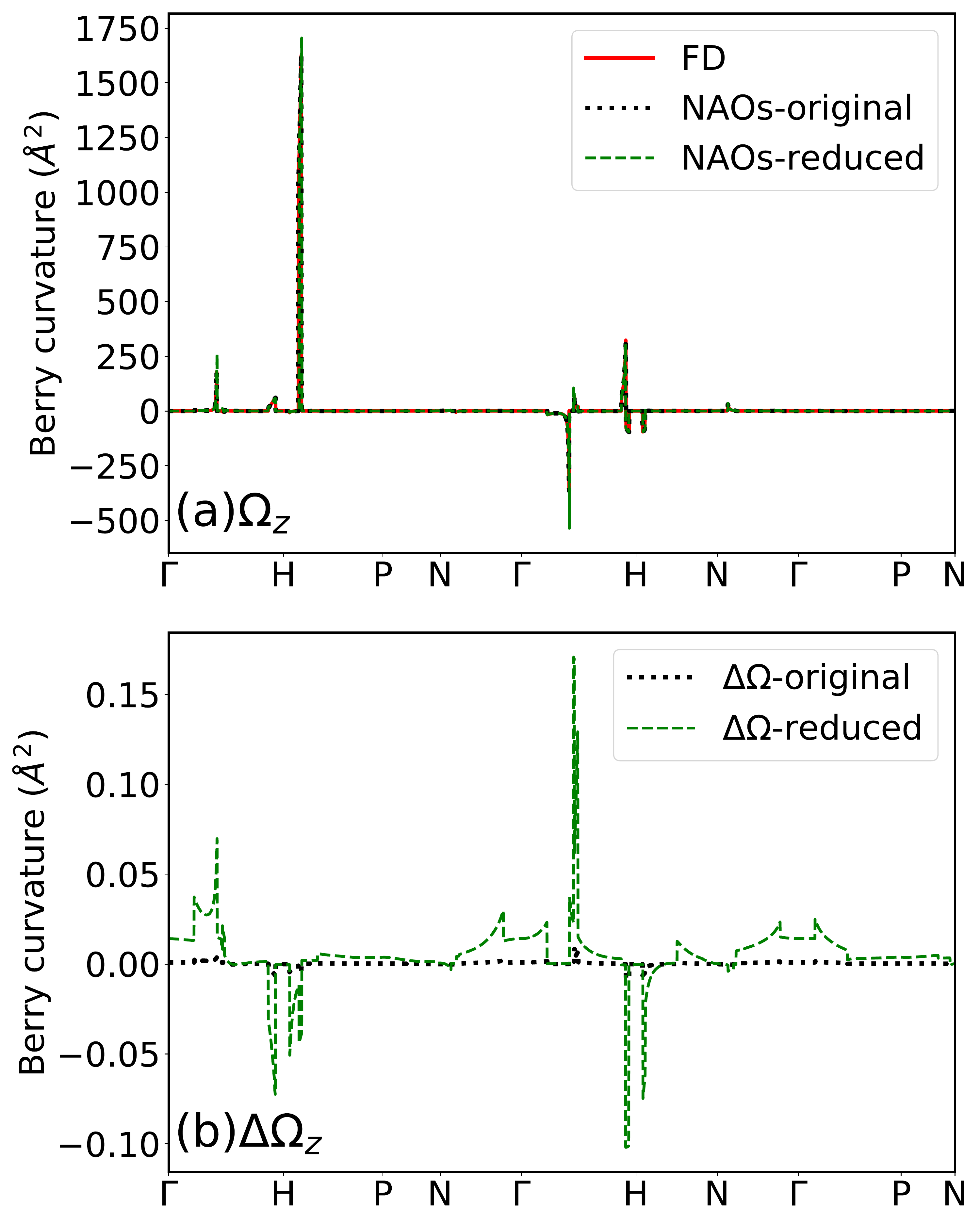}
	\caption{(a) Berry curvature $\Omega_z$ of bcc Fe along the high symmetry $k$ points. (b) The corrections $\Delta \Omega_z$ to the Kubo formula. The solid red line is calculated by FD, and the black dotted lines are  results of the original bases, whereas the green dashed lines are the results of the reduced bases. }
	\label{fig:fig3}
\end{figure}

\subsection{AHC of Fe}

In this section, we calculate the Berry curvature and the AHC for bcc Fe.
The lattice constant of bcc Fe is taken as $a$ = 2.870 \AA.
The band structure of Fe is shown in Fig. \ref{fig:bands}(b) in red solid lines around Fermi level $E_F$=0.

The Berry curvature of bcc Fe along the $c$-axis is given in Fig.~\ref{fig:fig3}(a). The black dotted line is the Berry curvature calculated via Eq.~(\ref{eq:total_berrycurvature}), which is in excellent agreement with the Berry curvature calculated via FD method, shown
in the red solid line. For most $k$ points, the Berry curvature is very small except at a few $k$ points, which have huge
Berry curvature, due to the spin-orbit-split avoided crossings near the Fermi level~\cite{Yao2004,Wang2006}.

We then calculate the dc anomalous Hall conductivity (AHC) \cite{Sundaram1999,Adams1959},
\begin{equation}
\sigma_{xy} = -\frac{e^2}{\hbar}\sum_{n}\int_{BZ}\frac{d{\bf k}}{(2\pi)^3}f_n({\bf k})\Omega_{n,z}({\bf k})\, .
\end{equation}
To calculate the AHC of bcc Fe, a 300$\times$300$\times$300 $k$-mesh is used and if the Berry curvature of a certain $k$ point is greater than 100 \AA$^{2}$, the Berry curvature is recalculated at a refined 7$\times$7$\times$7 submesh around the
$k$ point, following the scheme of Ref.~\cite{Yao2004,Wang2006}. The calculated AHC for Fe is 738 ($\Omega$ cm)$^{-1}$.
This result is in good agreement with 751 ($\Omega$cm)$^{-1}$, obtained
from full-potential all-electron calculations by Yao et. al.~\cite{Yao2004}, which is also close
to 756 ($\Omega$ cm)$^{-1}$ obtained from normal conserving pseudopotential
calculations via Wannier interpolation techniques~\cite{Wang2006}.
The Fe pseudopotential in Ref.~\cite{Wang2006} is specially optimized to reproduce the all-electron result of AHC,
and the difference between this work and Ref.~\cite{Wang2006}
might come from the different pseudopotentials used in the calculations.

\section{Berry curvatures calculated with reduced basis sets}
\label{sec:reduce_NAO}

In the previous section, we showed that the Berry curvatures calculated by Eq.~(\ref{eq:total_berrycurvature}) are in excellent agreement with FD results.
However, the NAO basis sets have too many orbitals compared to the Wannier bases, and therefore computationally more expensive. We would like to reduce the basis size to accelerate the calculations, but maintain the accuracy.
The idea is to use reduced basis sets, to reproduce the band structures in a smaller energy window. This is done via a revised band interpolation technique \cite{Chen2011} via NAOs.

To reduce the number of NAOs,  we reconstruct a new set of NAOs as the linear combination of original on-site orbitals, i.e,
\begin{equation}
|\tilde{\phi}_{I,\mu}\rangle = \sum_\nu^{N_I} \mathcal{U}_{I,\nu\mu}|\phi_{I,\nu}\rangle\, ,
\label{eq:redcued_NAOs}
\end{equation}
where $|\phi_{I,\nu}\rangle$ is the $\nu$-th orbit of the $I$-th atom.
One of the problems of the maximally localized Wannier functions is that their centers are not necessarily on atom positions or other high-symmetry points\cite{Sakuma2013}. In contrast, the reduced NAOs $|\tilde{\phi}_{I,\mu} \rangle$ are linear combinations of on-site NAOs, so they are still atom centered and strictly localized.
The number $\tilde{N}_I$ of the reduced NAO bases $|\tilde{\phi}_{I,\mu}\rangle$, is less than the number $N_I$ of $|\phi_{I,\nu}\rangle$. $\mathcal{U}_{I}$ is a real orthogonal matrix to keep the reduced NAOs real.

We obtained the reduced NAOs by minimizing the {\it spillage} between the wave functions calculated by original NAOs and the reduced basis set at a coarse $k$-mesh in a chosen energy window. Details of this process are given in \ref{sec:reduce_bases}.

\subsubsection{BaTiO$_3$}

Figure \ref{fig:bands}(a) compares the band structure of BaTiO$_3$ calculated by the original bases (solid red lines) and the reduced bases (black dashed lines).
The reduced bases set is optimized by fitting the wave functions in the energy window of -6 to 4.8 eV (The energy window is represented by a solid green line) which covers the lowest three bands above the Fermi level.
A uniform grid of 8$\times$8$\times$8 $k$-mesh is used to generate the reduced basis set.
The reduced base has 38 orbitals compared to the 86 orbitals in the original bases.
As we see from Fig.~\ref{fig:bands}(a), the band structures calculated by the reduced bases
are in excellent agreement with the original band structures within the energy window.
The lower energy bands above the energy window also agree very well.
For the bands far above the energy window, the agreement becomes worse as expected.

The Berry curvatures of BaTiO$_3$ calculated by the reduced bases are shown in green dashed lines in Fig.~\ref{fig:fig2}(a)-(c), compared to those calculated by the original bases. The overall agreement is rather good, with only small differences at some $k$-point.

We now check the $\Delta \Omega$ for the reduced bases, depicted in Fig.~~\ref{fig:fig2}(d)-(e), which are shown in green dashed lines. We see that for the reduced basis, $\Delta \Omega$ is significantly larger than that of the full basis calculations, which suggests that if a reduced basis is used, the corrections to the Kubo formula may not be ignored in some cases.
One may expect that if an even smaller non-orthogonal Wannier bases are used, the correction may have an even larger contribution, which cannot be ignored.

\subsubsection{bcc Fe}

The reduced basis set of the Fig. \ref{fig:bands}(b) is optimized in the energy window of -9 $\sim$ 5 eV, on a uniform grid of 8$\times$8$\times$8 $k$-points. We obtain 28 orbitals (include spin) from the original 54 orbitals. The energy bands obtained by the reduced basis set are almost identical to the original bands.
In fact, the agreement is still very well even outside the energy window, below 14 eV.

The Berry curvature $\Omega_z$ calculated by the reduced bases are also shown in Fig.~\ref{fig:fig3}(a) in the green dashed line compared to the original result. There are only very small differences between the results obtained by the original and reduced bases.
The correction to the naive Kubo formula $\Delta \Omega_z$ for the reduced bases are shown in the green dashed line
in  Fig.~\ref{fig:fig3}(b). The correction is much larger than that for the original bases, shown in the black dotted line. However, the correction is still negligibly small compared to the total Berry curvature, because the dominant contribution to the contribution comes from the $D$-$D$ terms~\cite{Wang2006}.

The AHC of bcc Fe calculated by the reduced basis set is 734 ($\Omega$ cm)$^{-1}$, which is in good agreement with that calculated by the full basis set 738 ($\Omega$ cm)$^{-1}$.

\section{Summary}
\label{sec:summary}

We derive the formula to calculate the Berry curvature using non-orthogonal NAOs. We find that there are some additional correction terms besides the usual Kubo formula. We calculate the Berry curvature of Rhombohedral BaTiO$_3$ and bcc Fe, as well as the AHC for Fe. The results are in excellent agreement with the finite difference method.
We develop a method that can significantly reduce the number of orbitals in the NAO bases but can still maintain the accuracy of the calculations. We also compare the Berry curvature and AHC calculated via our formula and the Kubo formula.
We find for the original basis, the differences between the two methods are negligibly small, but for the reduced bases sets, the correction terms become larger. The methods developed in this work can be applied to non-orthogonal generalized Wannier functions.


\appendix

\section{Electronic structure by LCAO}

In the NAO basis, the Hamiltonian matrix is calculated as,
\begin{equation}
H_{\nu\mu}(\mathbf{k}) = \sum_{\mathbf{R}}\mathrm{e}^{i\mathbf{k}
\cdot\mathbf{R}}\langle\mathbf{0}\nu|\hat{H}|\mathbf{R}\mu\rangle \, ,
\end{equation}
where $|\mathbf{R}\mu\rangle$ is a short notation of the $\mu$-th NAO in the ${\bf R}$-th unit cell, i.e.,
$\phi_{\mu}({\bf r}- {\bf R} -\tau_{\mu})$, and $\tau_{\mu}$ is the center of the orbital.
The Hamiltonian $\hat{H}$ is self-consistently determined by the Kohn-Sham equations.
In general, the NAOs are not orthogonal to each other, and their overlap matrix is
\begin{equation}
S_{\nu\mu}(\mathbf{k}) = \sum_{\mathbf{R}}\mathrm{e}^{i\mathbf{k}
\cdot\mathbf{R}}\langle\mathbf{0}\nu|\mathbf{R}\mu\rangle\, .
\end{equation}
Therefore, in the NAO bases, the corresponding Kohn-Sham equation (under
the converged charge density) is
\begin{equation}\label{hamilton_k}
H(\mathbf{k})C_n(\mathbf{k}) = E_{n\mathbf{k}}S(\mathbf{k})C_n(\mathbf{k})\, .
\end{equation}
The Bloch wave function is,
\begin{equation}
|\Psi_{n\mathbf{k}}\rangle = \frac{1}{\sqrt{N}}\sum_{\mu}C_{n\mu}(\mathbf{k})
\sum_{\mathbf{R}}\mathrm{e}^{i\mathbf{k}\cdot\mathbf{R}}|\mathbf{R}\mu\rangle\, ,
\end{equation}
and the corresponding cell-periodic part is
\begin{equation}
|u_{n\mathbf{k}}\rangle = \frac{1}{\sqrt{N}}\sum_{\mu}C_{n\mu}(\mathbf{k})\sum_{\mathbf{R}}\mathrm{e}^{i\mathbf{ k}\cdot(\mathbf{R}-\mathbf{r})}|\mathbf{R}\mu\rangle \, .
\label{eq:u_eq}
\end{equation}

\subsection{Linear response theory in non-orthogonal NAO bases}

According to the perturbation theory, we derive the first-order term of Eq.~(\ref{hamilton_k}) with respect to the parameter $\mathbf{k}$ in the non-degenerate case,
\begin{eqnarray}
&\left[ E_{n}S - H \right]\left(\partial_\alpha C_{n}\right) \nonumber\\
&= \left[ \partial_\alpha H - E_{n}\left(\partial_\alpha S\right) - \left(\partial_\alpha E_{n}\right)S \right]C_{n} \label{perturbation}
\end{eqnarray}
We use $\{C_{m}\}$ set to expand $\partial_\alpha C_{n}$, i.e.,
\begin{equation}
\partial_\alpha C_{n} = \sum_{m} \left[C_{m}^\dagger S\left(\partial_\alpha C_{n}\right)\right] C_{m} \, .
\end{equation}
Multiply $C_{m}^\dagger$ to the left of Eq.~(\ref{perturbation}), we have
\begin{eqnarray}
& (E_{n} - E_{m})C_{m}^\dagger S\left(\partial_\alpha C_{n}\right) \nonumber \\
&= C_{m}^\dagger\left[ \partial_\alpha H - E_{n}\left(\partial_\alpha S\right)\right]C_{n} \, .
\end{eqnarray}
We obtain for $m\neq n$,
\begin{equation}
C_{m}^{\dagger}S\left(\partial_\alpha C_{n}\right) = \frac{C_{m}^{\dagger}\left(\partial_\alpha H\right)C_n - E_{n}C_m^\dagger\left(\partial_\alpha S\right)C_{n}}{E_{n} - E_{m}} \, .
\end{equation}
Generally, for the non-orthogonal bases,
\begin{equation}
C_{n}^{\dagger}S \left(\partial_\alpha C_{n}\right) \neq 0 \, ,
\end{equation}
and can be determined by choice of a particular gauge as discussed in Sec.~\ref{sec:IIB}.


\subsection{Relations between $\mathbf{\bar{A}}$ and $\mathbf{\bar{A}}^\dagger$, $\mathbf{D}$ and $\mathbf{D}^\dagger$}

There are two useful relations between
$\bar{{\bf A}}$ with $\bar{{\bf A}}^\dagger$ and ${\bf D}$ with ${\bf D}^\dagger$  that appear in Eq.~(\ref{eq:relation1}) and Eq.~(\ref{eq:relation2}), respectively.

To prove Eq.~(\ref{eq:relation1}), we note that,
the conjugated form of Eq.~(\ref{eq:dipole}),
\begin{eqnarray}
(A^{R\dagger})_{\mu\nu,\alpha}
&=& \sum_{\bf R}\mathrm{e}^{-i{\bf k}\cdot {\bf R}}\langle{\bf R}\mu|r_\alpha|{\bf 0}\nu\rangle \nonumber \\
&=& \sum_{\bf R}\mathrm{e}^{-i{\bf k}\cdot {\bf R}}\langle{\bf 0}\mu|r_\alpha + R_\alpha|{\bf -R}\nu\rangle \nonumber \\
&=& A^{R}_{\mu\nu,\alpha} + i\left(\partial_\alpha S\right)_{\mu\nu}
\end{eqnarray}
We can readily get the relation,
\begin{equation}
\bar{A}_{nm,\alpha} - (\bar{A}^\dagger)_{nm,\alpha} = -i\bar{S}_{nm,\alpha}\, .
\end{equation}
Due to the orthonormal condition,
\begin{equation}
C_n^\dagger({\bf k})S({\bf k}) C_m({\bf k}=\delta_{mn}\, ,
\end{equation}
at each ${\bf k}$ point, we have
$\partial_{\alpha} \left[C_n^\dagger ({\bf k})S({\bf k}) C_m({\bf k})\right] =0$,
i.e.,
\begin{equation}
\left(\partial_{\alpha} C_n^\dagger\right) S C_m
 + C_n^\dagger \left(\partial_{\alpha} S\right) C_m + C_n^\dagger S \left(\partial_{\alpha}  C_m\right) = 0 \,.
\label{eq:orthnormal}
\end{equation}
which gives the relation Eq.~(\ref{eq:relation2}),
\begin{equation}
D_{nm,\alpha} + (D^\dagger)_{nm,\alpha} = -\bar{S}_{nm,\alpha}\, .
\end{equation}

\subsection{Proof of gauge invariance of Berry curvature}
\label{sec:berry_curvature_gauge}

We prove the Berry curvature $\Omega_{\alpha\beta}$ defined in Eq.~(\ref{eq:modify_omega})
is gauge invariant. When $C_n$ is changed to $\tilde{C}_n=\mathrm{e}^{i\xi_n}C_n $,
the corresponding $\mathbf{D}$, $\mathbf{\bar{S}}$,$\mathbf{\bar{A}^\dagger}$ matrix will change to,
\begin{eqnarray}
\tilde{D}_{nm,\alpha} &=& \mathrm{e}^{i(\xi_m-\xi_n)}\left(D_{nm,\alpha} + i\xi_m^\prime\delta_{nm}\right) , \nonumber \\
\tilde{\bar{S}}_{nm,\alpha} &=& \mathrm{e}^{i(\xi_m-\xi_n)}\bar{S}_{nm,\alpha} \nonumber , \\
\left(\tilde{\bar{A}}^\dagger\right)_{nm,\alpha} &=& \mathrm{e}^{i(\xi_m-\xi_n)}\left(\bar{A}^\dagger\right)_{nm,\alpha}.
\end{eqnarray}
The fourth term in Eq.~(\ref{eq:modify_omega}) becomes
\begin{eqnarray}
\left[\tilde{D}_\alpha,\tilde{D}_\beta\right]_{nn}
&=& \sum_{m} \tilde{D}_{nm,\alpha}\tilde{D}_{mn,\beta} - \tilde{D}_{nm,\beta}\tilde{D}_{mn,\alpha} \nonumber \\
&=& \sum_{m}
\left(D_{nm,\alpha}+i\xi_{m}^\prime\delta_{nm}\right)
\left(D_{mn,\beta}+i\xi_{n}^\prime\delta_{mn}\right) \nonumber \\
&-& \sum_{m}
\left(D_{nm,\beta}+i\xi_{m}^\prime\delta_{nm}\right)
\left(D_{mn,\alpha}+i\xi_{n}^\prime\delta_{mn}\right) \nonumber \\
&=& \sum_{m} {D}_{nm,\alpha}{D}_{mn,\beta} - {D}_{nm,\beta}{D}_{mn,\alpha} \nonumber\\
&=& \left[D_\alpha,D_\beta\right]_{nn} ,
\end{eqnarray}
i.e., it is invariant under the gauge change.
Similarly, we can prove that all other terms of Eq.~(\ref{eq:modify_omega}) are
also gauge invariant, and therefore the Berry curvature is gauge invariant.

\subsection{Calculating the Berry curvature by finite difference}

The Berry curvature can be calculated via the finite difference method,
using the relationship between Berry phase and Berry curvature \cite{vanderbilt_2018},
\begin{equation}
\Delta \phi=\oint A({\bf k}) dl_{\bf k}= \int_{S} {\bf \Omega}({\bf k}) \cdot d{\bf S} \, .
\label{eq:FD}
\end{equation}
We choose a small square loop around a ${\bf k}$ point in the Brillouin Zone and calculate
the discretized Berry phase \cite{King-Smith1993} of the loop.
The Berry curvature at the given ${\bf k}$ point can be obtained as ${\bf \Omega} ({\bf k}) \cdot \hat{\bf n}$=$\Delta \phi / \Delta S$, where $\hat{\bf n}$ is the normal direction of the square loop and $\Delta S$ is its area.
In principles the Berry phase is correct only modulo 2$\pi$. \cite{vanderbilt_2018}
But here, we have no ambiguity of chosen the correct branches,
as $\Delta \phi$ is significantly smaller than the quanta  when $\Delta S$ is very small.

\section{Optimize the reduced bases}
\label{sec:reduce_bases}

The reduced bases orbitals $ \{ \tilde{\phi}_{I,\nu} \}$ are obtained by minimizing the {\it spillage} between the reduced bases and the Bloch wave functions in the selected energy window and $k$ points. The spillage \cite{Chen2011} is defined as,
\begin{equation}
\mathcal{S} = \frac{1}{N_n N_k}\sum_{n}\sum_{\mathbf{k}}
\langle\Psi_{n\mathbf{k}}|1-\hat{P}(\mathbf{k})|\Psi_{n\mathbf{k}}\rangle
\end{equation}
where $n$ is the energy bands within the energy window.
$\hat{P}(\mathbf{k})$ is a projector spanned by the reduced NAOs,
\begin{equation}
\hat{P}(\mathbf{k}) = \sum_{\mu,\nu}|\mathbf{k},\tilde{\phi}_{\mu}\rangle \tilde{S}_{\mu\nu}^{-1}(\mathbf{k})\langle\mathbf{k},\tilde{\phi}_{\nu}| \, ,
\end{equation}
where, $\tilde{S}_{\mu\nu}(\mathbf{k})$ is the overlap matrix of the reduced NAOs.
The reduced NAOs are the linear combination of the on-site original NAOs.
Using Eq.~(\ref{eq:redcued_NAOs}), we have,
\begin{equation}
\tilde{S}_{\mu\nu}(\mathbf{k}) = \sum_{\alpha,\beta}\mathcal{U}_{\alpha\mu}^* S_{\alpha\beta}(\mathbf{k})\mathcal{U}_{\beta\nu}
= \left(\mathcal{U}^\dagger S(\mathbf{k})\mathcal{U}\right)_{\mu\nu} \, ,
\end{equation}
where $S(\mathbf{k})$ is the overlap matrix of the original NAOs.
The spillage can be written as,
\begin{equation}
\mathcal{S}=\frac{1}{N_n N_k}\sum_{n}\sum_{\mathbf{k}}
\left[1 -
C_n^\dagger S \mathcal{U}\left(\mathcal{U}^\dagger S \mathcal{U}\right)^{-1}\mathcal{U}^\dagger S C_n
\right]\, .
\end{equation}

Using the rule,
\begin{equation}
\mathrm{d}(X^{-1}) = -X^{-1}\left[\mathrm{d}X\right]X^{-1} \,
\end{equation}
where $X$ is an invertible matrix,
we get the derivative of $Spillage$ to the $\mathcal{U}$ matrix,
\begin{eqnarray}
& \frac{\partial \mathcal{S}}{\partial \mathcal{U}} = \frac{1}{N_n N_k}\sum_{n,\mathbf{k}}
\left[
-\left(
\left(\mathcal{U}^\dagger S \mathcal{U}\right)^{-1}\mathcal{U}^\dagger SC_nC_n^\dagger S
\right)^T \right. \nonumber \\
&+
S\mathcal{U}\left(\mathcal{U}^\dagger S\mathcal{U}\right)^{-1}\mathcal{U}^\dagger SC_nC_n^\dagger S\mathcal{U}\left(\mathcal{U}^\dagger S\mathcal{U}\right)^{-1} \nonumber \\
&+ \left(
\left(\mathcal{U}^\dagger S\mathcal{U}\right)^{-1}\mathcal{U}^\dagger SC_nC_n^\dagger S\mathcal{U}\left(\mathcal{U}^\dagger S\mathcal{U}\right)^{-1}\mathcal{U}^\dagger S
\right)^T \nonumber \\
&\left.
- SC_nC_n^\dagger S\mathcal{U}\left(\mathcal{U}^\dagger S\mathcal{U}\right)^{-1} \right]
\label{eq:gradients}
\end{eqnarray}
Once we have the gradients, we optimize the spillage via the Broyden–Fletcher–Goldfarb–Shanno (BFGS) method implemented in the numpy~\cite{Harris2020_numpy} and scipy package.\cite{Virtanen2020_scipy}
To determine the initial reduced NAOs, we evaluate the coefficients of each orbital in the full bases of the wave functions
in the selected energy window and $k$-mesh. We keep the NAOs, whose coefficients are larger than some threshold as the initial reduced NAOs, which give the corresponding initial $\mathcal{U}$ matrices. We then minimize the spillage to obtain the optimized NAOs. When optimize the $\mathcal{U}$ matrix using Eq.~(\ref{eq:gradients}), the wave functions $C_n$, overlap matrix $S$, and initial $\mathcal{U}$ matrix all have the point group symmetry, and therefore in principle, the reduced NAOs should still preserve the symmetry. However, there might be small symmetry breaking, due to numerical noise. The reduced NAOs can be symmetrized using the algorithms of Ref.~\cite{Sakuma2013} if necessary.

\section*{References}

\end{document}